\documentstyle{mn}

\newif\ifAMStwofonts

\ifoldfss
  \ifCUPmtlplainloaded \else
    \NewTextAlphabet{textbfit} {cmbxti10} {}
    \NewTextAlphabet{textbfss} {cmssbx10} {}
    \NewMathAlphabet{mathbfit} {cmbxti10} {} 
    \NewMathAlphabet{mathbfss} {cmssbx10} {} 
  \fi
  \ifAMStwofonts
    \ifCUPmtlplainloaded \else
      \NewSymbolFont{upmath} {eurm10}
      \NewSymbolFont{AMSa} {msam10}
      \NewMathSymbol{\upi}     {0}{upmath}{19}
      \NewMathSymbol{\umu}     {0}{upmath}{16}
      \NewMathSymbol{\upartial}{0}{upmath}{40}
      \NewMathSymbol{\leqslant}{3}{AMSa}{36}
      \NewMathSymbol{\geqslant}{3}{AMSa}{3E}

       \let\ge=\geqslant
    \fi
  \fi
\fi 

\ifnfssone
  \newmathalphabet{\mathit}
  \addtoversion{normal}{\mathit}{cmr}{m}{it}
  \addtoversion{bold}{\mathit}{cmr}{bx}{it}
  \newmathalphabet{\mathbfit} 
  \addtoversion{normal}{\mathbfit}{cmr}{bx}{it}
  \addtoversion{bold}{\mathbfit}{cmr}{bx}{it}
  \newmathalphabet{\mathbfss} 
  \addtoversion{normal}{\mathbfss}{cmss}{bx}{n}
  \addtoversion{bold}{\mathbfss}{cmss}{bx}{n}
  \ifAMStwofonts
    \ifCUPmtlplainloaded \else
      \UseAMStwoboldmath
      \makeatletter
      \new@mathgroup\upmath@group
      \define@mathgroup\mv@normal\upmath@group{eur}{m}{n}
      \define@mathgroup\mv@bold\upmath@group{eur}{b}{n}
      \edef\UPM{\hexnumber\upmath@group}
      \new@mathgroup\amsa@group
      \define@mathgroup\mv@normal\amsa@group{msa}{m}{n}
      \define@mathgroup\mv@bold\amsa@group{msa}{m}{n}
      \edef\AMSa{\hexnumber\amsa@group}
      \makeatother
      \mathchardef\upi="0\UPM19
      \mathchardef\umu="0\UPM16
      \mathchardef\upartial="0\UPM40
      \mathchardef\leqslant="3\AMSa36
      \mathchardef\geqslant="3\AMSa3E

       \let\ge=\geqslant
    \fi
  \fi
\fi 

\ifnfsstwo
  \DeclareMathAlphabet{\mathbfit}{OT1}{cmr}{bx}{it}
  \SetMathAlphabet\mathbfit{bold}{OT1}{cmr}{bx}{it}
  \DeclareMathAlphabet{\mathbfss}{OT1}{cmss}{bx}{n}
  \SetMathAlphabet\mathbfss{bold}{OT1}{cmss}{bx}{n}
  \ifAMStwofonts
    \ifCUPmtlplainloaded \else
      \DeclareSymbolFont{UPM}{U}{eur}{m}{n}
      \SetSymbolFont{UPM}{bold}{U}{eur}{b}{n}
      \DeclareSymbolFont{AMSa}{U}{msa}{m}{n}
      \DeclareMathSymbol{\upi}{0}{UPM}{"19}
      \DeclareMathSymbol{\umu}{0}{UPM}{"16}
      \DeclareMathSymbol{\upartial}{0}{UPM}{"40}
      \DeclareMathSymbol{\leqslant}{3}{AMSa}{"36}
      \DeclareMathSymbol{\geqslant}{3}{AMSa}{"3E}

       \let\ge=\geqslant
    \fi
  \fi
\fi 

\ifCUPmtlplainloaded \else
  \ifAMStwofonts \else 
    \def\upi{\pi}
    \def\umu{\mu}
    \def\upartial{\partial}
  \fi
\fi

\title{Gamma-ray and neutrino flares produced by protons 
accelerated on an accretion disk surface in AGN}
\author[W. Bednarek and R.J. Protheroe]
{W. Bednarek$^{1,2}$ and R.J. Protheroe$^{1}$ \\
$^1$Department of Physics and Mathematical Physics,
The University of Adelaide, Adelaide, Australia 5005.\\
$^2$Department of Experimental Physics, University of \L\'od\'z,
90-236 \L\'od\'z, ul. Pomorska 149/153, Poland.}

\date{Accepted 199 Received 199; in original form 1998}

\pubyear{199}

\begin{document}

\maketitle

\label{firstpage}

\begin{abstract}

We discuss the consequences of almost rectilinear acceleration of
protons to extremely high energies in a reconnection region on
the surface of an accretion disk which surrounds a central black
hole in an active galaxy. The protons produce $\gamma$-rays and
neutrinos in interactions with the disk radiation as considered
in several previous papers. However, in this model the secondary
$\gamma$-rays can initiate cascades in the magnetic and radiation
fields above the disk. We compute the spectra of $\gamma$-rays
and neutrinos emerging from regions close to the disk surface.
Depending on the parameters of the reconnection regions, this
model predicts the appearance of $\gamma$-ray and neutrino flares
if protons takes most of the energy from the reconnection
region. In contrast, if leptons take most of the energy, they
produce pure $\gamma$-ray flares. The $\gamma$-ray spectrum
expected in the case of hadronic cascading is compared with the
spectrum observed during the flare in June 1991 from 3C 279.  The
neutrino flares which should accompany these gamma-ray flares may
be detected by future large scale neutrino telescopes sensitive
at $\sim 10^{5}$ TeV.

\end{abstract}


\section{Introduction}

Flares on very short time scales in TeV $\gamma$-rays have been
recently detected from Mrk 421, Mrk 501, and 1ES2344+514
(e.g. Buckley et al.~1996, Gaidos et al.~1996, Aharonian et al.~
1997, Catanese et al.~1998) and in the EGRET energy range from
many other blazars (e.g. von Montigny et al.~1995, Mattox et
al.~1997).  Flares observed with a $\sim 1$~day time scale are
usually interpreted in terms of a relativistic shock moving
through the jet with a Lorentz factor of the order of $\sim 10$
(e.g. Maraschi, Ghisellini \& Celloti~1992, Mannheim \&
Biermann~1992, Dermer \& Schlickeiser~1993, Bednarek~1993,
Sikora, Begelman \& Rees~1994, Blandford \& Levinson~1995,
Ghisellini \& Madau~1996, Protheroe~1997).  However, the
production of such flares may also be a natural consequence of
energization of particles in a way similar to that causing solar
flares, but in this case occurring in regions of magnetic
reconnection which may exist on the disk surface (Haswell, Tajima
\& Sakai~1992, Lesch \& Pohl~1992).  Reconnection of magnetic
field may also occur in the jet (e.g. Romanova \& Lovelace~1992,
Bisnovatyi-Kogan \& Lovelace~1995). The processes of $\gamma$-ray
production by leptons accelerated in electric fields generated in
the jet has been recently discussed by Bednarek \& Kirk (1995,
henceforth BK95) and Bednarek, Kirk \& Mastichiadis (1996,
henceforth BKM96).

The acceleration of particles by almost rectilinear electric
fields has some advantages in comparison to the stochastic
acceleration by a relativistic shock since, in principle,
particles can reach very high energies on shorter time
scales. However, the process of reconnection of magnetic fields
in astrophysical environments is poorly understood. Recently,
Haswell, Tajima \& Sakai (1992) have developed a model of
explosive reconnection of magnetic fields on the surface of an
accretion disk.  They argued that particles (electrons, protons)
can be accelerated to extremely high energies ($\sim 10^{20-21}$
eV).  The consequences of this model for $\gamma$-ray production
in blazars has been investigated recently by Bednarek (1997,
henceforth B97), assuming that accelerated leptons develop
cascades in the magnetic field through production of
$\gamma$-rays by the synchrotron process, and by the creation of
secondary pairs by $\gamma$-rays in the magnetic field. The
$\gamma$-ray spectra, emerging from a region on the inner part of
an accretion disk around a massive black hole, are in agreement
with observations of blazars.
 
In this paper, we discuss the consequences of acceleration of
protons in almost uniform electric fields induced by magnetic
reconnection on the disk surface. Protons accelerated to
extremely high energies produce $\gamma$-rays and neutrinos
through the decay of pions produced in collisions with disk
photons. Neutrinos freely escape, but $\gamma$-ray photons
initiate cascades in the magnetic and radiation fields above the
disk (B97).  The conditions for acceleration of protons,
and their propagation in the disk radiation, are discussed in
Sect.~2.1 and 2.2.  We compute the spectra of neutrinos and
$\gamma$-rays (Sect.~2.3), and describe the details of the
cascades initiated by these $\gamma$-rays (Sect.~3). Example
$\gamma$-ray spectra are compared with observations of the June
1991 flare from 3C 279, and a prediction of the neutrino emission
accompanying this flare is given in Sect.~4.

\section{Acceleration of protons during explosive reconnection in 
the inner part of an accretion disk}

We make use of the model of Haswell et al.~(1992) in which
particles can be accelerated to extremely high energies during
explosive reconnection of the magnetic field on the surface of an
accretion disk in the central parts of active galactic nuclei
(AGN). Let us assume that reconnection of the magnetic field
occurs over a distance $l_{\rm rec}$ somewhere in the inner part
of the optically thick accretion disk. The local radiation field
in this region can be described by a black body spectrum with
temperature $T$. The induced electric field strength $E_{\rm f}$ can be
estimated as $E_{\rm f}\cong c B$, where $B$ is the reconnecting
magnetic field strength and $c$ is the velocity of light.  Then,
the maximum possible energies, which can be reached by particles
(in eV), are
\begin{eqnarray}
E_p = E_{\rm f} l_{\rm rec}\approx 3\times 10^{18} B_5 l_{11},
\label{eq1}
\end{eqnarray}
\noindent
where $B = 10^5B_5$ G, $l_{\rm rec} = 10^{11}l_{11}$ cm.

\subsection{Conditions for acceleration of protons}

Protons accelerated in the electric field lose energy by
$e^{\pm}$ pair and pion production in collisions with the black
body photons after reaching the thresholds for these reactions.
Since the threshold for $e^{\pm}$ pair production is about two
orders of magnitude lower than that for pion production, $e^\pm$
pairs are first injected into the electric field in the
reconnection region.  The characteristic distance scale
$\lambda_{{\rm p}\gamma e^{\pm}}$ (in cm), for this process to
occur efficiently, can be estimated by
\begin{eqnarray}
\lambda_{{\rm p}\gamma e^{\pm}}\approx 
(\sigma_{{\rm p}\gamma\rightarrow e^{\pm}}
n_{\rm bb})^{-1}\approx 5.1\times 10^{12}T_4^{-3}, 
\label{eq2}
\end{eqnarray}
\noindent
where $\sigma_{{\rm p}\gamma\rightarrow e^{\pm}}\approx 10^{-26}$
cm$^2$ is the cross section for $e^{\pm}$ pair production in
collisions of protons with photons (Maximon~1968, Chodorowski,
Zdziarski \& Sikora~1992), $n_{\rm bb}$ is density of the black
body photons, and the local disk temperature in the reconnection
region is expressed in units of $T_4=T/10^4$ K.  We shall show
later that this process is not an important energy loss mechanism
for accelerated protons since its inelasticity coefficient is
small ($\sim 2\times 10^{-3}$).  Nevertheless, the injected
secondary $e^{\pm}$ pairs can initiate inverse Compton $e^{\pm}$
pair cascades in the disk radiation and the electric field of the
reconnection region (BK95).  The products of such a cascade can
saturate the external electric field if the length of the
reconnection region is longer than (BK95)
\begin{eqnarray}
\lambda_{\rm sat}\approx 100\lambda_{\rm trip}\approx 3.85\times 10^{14}
T_4^{-3},
\label{eq3}
\end{eqnarray}
\noindent
where $\lambda_{\rm trip}\approx (\sigma_{\rm trip}n_{\rm
bb})^{-1}$ is characteristic distance scale for the triplet
$e^{\pm}$ pair production by electrons in collision with soft
photons, and $\sigma_{\rm trip}\approx \sigma_{\rm T}/50$
(Mastichiadis~1991), $\sigma_{\rm T}$ is the Thomson cross
section.  In this case, most of the energy extracted from the
electric field is transferred to secondary $e^{\pm}$ pairs if
\begin{eqnarray}
l_{\rm rec} > \lambda_{\rm sat}. 
\label{eq3a}
\end{eqnarray}
\noindent
Eq.~\ref{eq3a} can be written as a condition on the temperature
of the black body radiation in the reconnection region such that
the electric field is saturated,
\begin{eqnarray}
T_4 > T_{4,{\rm tr}}\approx 15.7 l_{11}^{-1/3}.
\label{eq4}
\end{eqnarray}

If the temperature of the black body radiation is lower than
$T_{4,{\rm tr}}$ then the reconnection energy can efficiently
accelerate relativistic protons.  These protons can then produce
pions in interactions with the disk radiation both during the
acceleration stage and after emerging from the reconnection
region, provided that the electric field is strong enough to
allow them to reach energies, $E_{\rm p,min}$ (in eV), above the
threshold for pion production,
\begin{eqnarray}
E_{\rm p} > E_{\rm p,min} \cong 3\times 10^{16} T_4^{-1}.
\label{eq5}
\end{eqnarray}
\noindent
The characteristic distance scale for pion production by
protons can be estimated by,
\begin{eqnarray}
\lambda_{{\rm p}\gamma \pi}\approx (\sigma_{{\rm p}\gamma\pi}n_{bb})^{-1}
\approx 1.7\times 10^{14} T_4^{-3}
\label{eq6}
\end{eqnarray}
\noindent
where $\sigma_{{\rm p}\gamma \pi}\approx 3\times 10^{-28}$
cm$^{-2}$ is the cross section for pion production in collisions
of protons with photons (Stecker~1968).  Comparison of
$\lambda_{{\rm p}\gamma \pi}$ with the characteristic distance
scale of the reconnection region, which is the inner radius of an
accretion disk $r_{\rm in}$, allows us to estimate the parameters
of the accretion disk for which the proton energy losses by pion
production become significant.  We find that the minimum local
disk temperature in the reconnection region depends on the
characteristic disk dimension,
\begin{eqnarray}
T_4 > 1.2 r_{14}^{-1/3},
\label{eq6b}
\end{eqnarray}
\noindent
where $r_{\rm in} = 10^{14}r_{14}$ cm.

   \begin{figure}
      \vspace{7.cm}
\includegraphics{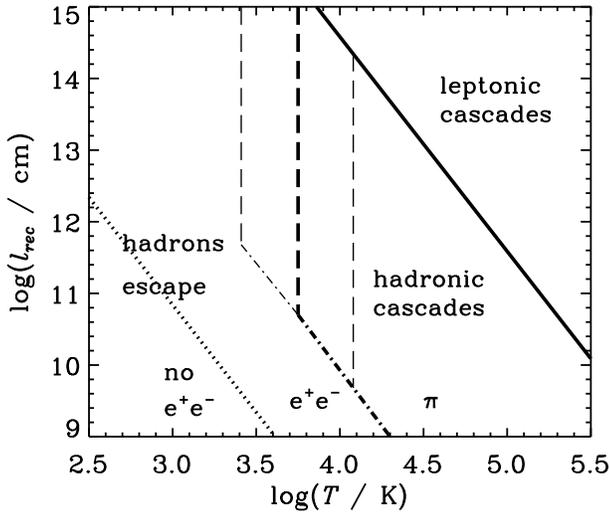}
      \caption[]{The conditions for cascading initiated in the
radiation field of the accretion disk by hadrons accelerated on
the disk surface.  For the parameters in the areas marked by
`hadrons escape', the protons escape from the radiation of the
disk without significant energy losses. In the area `no $e^+e^-$'
they have energies below the threshold for $e^{\pm}$ pair
production and in the area `$e^+e^-$', they inject $e^{\pm}$
pairs. For parameters in the area marked by `hadronic cascades',
protons efficiently lose energy on pion production and the pion
decay products initiate cascades in the magnetic and radiation
fields. In the area `leptonic cascades', the energy of the
electric field is extracted mainly by secondary leptons which
initiate cascades in the magnetic and radiation fields
(Eq.~5). The thick long-dashed line shows the limit for efficient
extraction of energy from hadrons (Eq.~8) for a disk inner
radius of $r_{\rm in} = 10^{15}$ cm, and the thin long-dashed lines
give the limits for $r_{\rm in} = 10^{14}$ cm (right) and
$10^{16}$ cm (left). }
\label{fig1}
    \end{figure}

The maximum proton energies, estimated using Eq.~1, can be
expressed in terms of the temperature of the black body radiation
in the reconnection region if we use the common assumption of
equipartition of the magnetic and radiation energy densities
close to the accretion disk. However, we note that Haswell,
Tajima \& Sakai~(1992) argue that the magnetic field in the
inner disk can be significantly amplified as a result of its
differential rotation from the values close to the equipartition
($B_{\rm eq}\approx 40 T_4^2$) up to values of the order of $B\approx
300 B_{\rm eq}$. In such a case, the formula for the proton energy
(Eq.~1) can be rewritten in the form
\begin{eqnarray}
E_{\rm p}\approx  3.6\times 10^{17}T_4^2 l_{11}.
\label{eq7}
\end{eqnarray}
\noindent
Since efficient proton acceleration requires the electric field
not to be saturated by pairs, the disk temperature in the
reconnection region cannot exceed that given by Eq.~\ref{eq4},
and we can express the maximum possible proton energy (in eV) in
terms of the local disk temperature
\begin{eqnarray}
E_{\rm p,max} \approx 1.4\times 10^{21} T_4^{-1}.
\label{eq8}
\end{eqnarray}
The assumption of equipartition between the primary magnetic
field energy density and the radiation energy density allows us
the derive the minimum length (in cm) of the reconnection region
for which accelerated protons can reach energies above the
thresholds for $e^{\pm}$ pair production and pion production
(Eq.~\ref{eq5}) giving
\begin{eqnarray}
l_{\rm rec}^{e^{\pm}} > 6.9\times 10^7 T_4^{-3},
\label{eq9}
\end{eqnarray}
\noindent
and
\begin{eqnarray}
l_{\rm rec}^{\pi} > 8.3\times 10^{9} T_4^{-3},
\label{eq10}
\end{eqnarray}
\noindent 
in cm, respectively.  The conditions for acceleration and
propagation of protons (Eqs.~5, 8, 11, and 12) allow us to
distinguish three interesting scenarios which can be separated in
the parameter space (shown in Fig.~\ref{fig1}) defined by the
length of the acceleration region $l_{\rm rec}$ and the
temperature of the local black body radiation $T$.  For the
regions labeled by `hadrons escape', the proton energy losses
during acceleration and subsequent propagation in the disk
radiation are negligible, allowing protons to leave the accretion
disk with very high energies. If the acceleration distance is
high enough, the protons can reach the threshold for $e^{\pm}$
pair production, injecting a few secondary pairs (area marked by
`$e^+e^-$', between the dot-dashed line (Eq.~\ref{eq10}), dotted
line (Eq.~\ref{eq9}) and long-dashed line
(Eq.~\ref{eq6b})). Below the dotted line, protons do not inject
pairs (area marked by `no $e^+e^-$'). If protons reach the
threshold for pion production (Eq.~\ref{eq10}, marked by the
dot-dashed line), they suffer strong energy losses. The proton
energy losses becomes significant for disk temperatures above the
values estimated by Eq.~\ref{eq6b} (marked by the long-dashed
lines for different disk inner radii). In the area marked by
`hadronic cascades', the initial proton's energy is transferred
to pions whose decay products (leptons, photons) initiate
cascades in the magnetic and radiation fields above the disk. In
the present paper, we shall concentrate on this case.  In the
last area, marked by `leptonic cascades', the production of
$e^\pm$ pairs by protons and subsequently by these pairs in the
inverse Compton $e^\pm$ pair cascade developing in the electric
field, is very efficient, even during the acceleration stage in
the reconnection region. These secondary $e^\pm$ pairs may even
saturate the electric field (BK95).  Most of the energy of the
electric field is then transferred to leptons (and $\gamma$-rays)
which initiate cascades in the magnetic and radiation fields
above the disk identical to those initiated by primary leptons
(B97). Note that the processes occurring for the parameters in
the area `hadronic cascades' turn out to be the production of
$\gamma$-ray and very high energy neutrino flares, whereas for
the processes occurring for parameters in the area `leptonic
cascades' only $\gamma$-ray flares are produced.

   \begin{figure}
      \vspace{9.cm}
\includegraphics{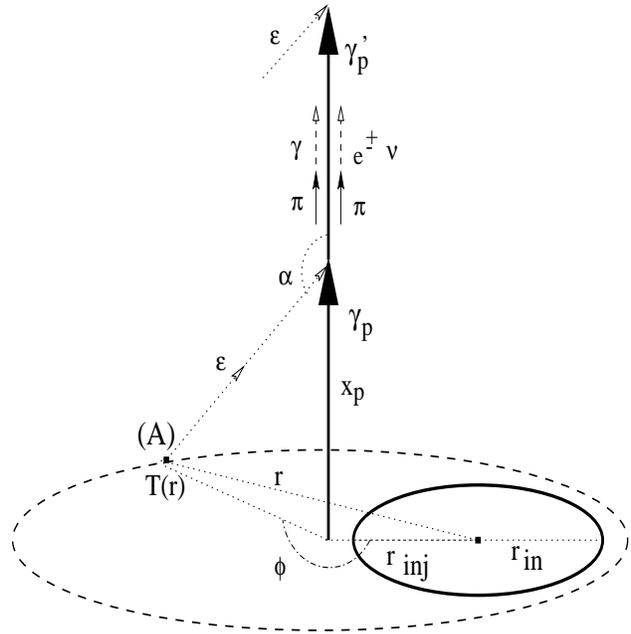}
      \caption[]{Schematic representation of the scenario
discussed in this paper. Relativistic protons with Lorentz
factors $\gamma_{\rm p}$ are injected at a reconnection region
which is at a distance $r_{\rm inj}$ from the center of the
disk. A proton interacts at height $x_{\rm p}$ with a soft photon
from the disk, having energy $\epsilon$, emitted at point A on
the disk defined by the angles $\alpha$ and $\phi$. The energy
$\epsilon$ is determined by the local disk temperature $T(r)$ at
point A. As a result, pions are produced which decay into
$\gamma$-rays, $e^{\pm}$ pairs, and neutrinos. The surviving
proton, with the Lorentz factor $\gamma_{\rm p'}$, may again
interact at some distance from the disk.}
\label{fig1b}
    \end{figure}

\subsection{Propagation of relativistic protons in the disk radiation}

The protons, escaping from the reconnection region with energies
estimated by Eq.~(\ref{eq7}) or (\ref{eq8}), produce pions in
collisions with the disk photons on a characteristic distance
scale given by Eq.~(\ref{eq6}).  We simulate propagation of
protons through the anisotropic disk radiation by using the Monte
Carlo method. It is assumed that protons are injected on the
surface of a thin disk at the distance $r_{\rm inj}$ from its
center (see Fig.~\ref{fig1b}).  We consider a simple disk model
in which the disk emits black body radiation with characteristic
temperature depending on the distance from the disk center
according to the model of Shakura \& Sunyaev~(1973).  Therefore,
the disk radiation is defined by the disk inner radius $r_{\rm
in}$ and its temperature $T_{\rm in}$ at $r_{\rm in}$.

\begin{figure*}
      \vspace{7.cm}
\includegraphics{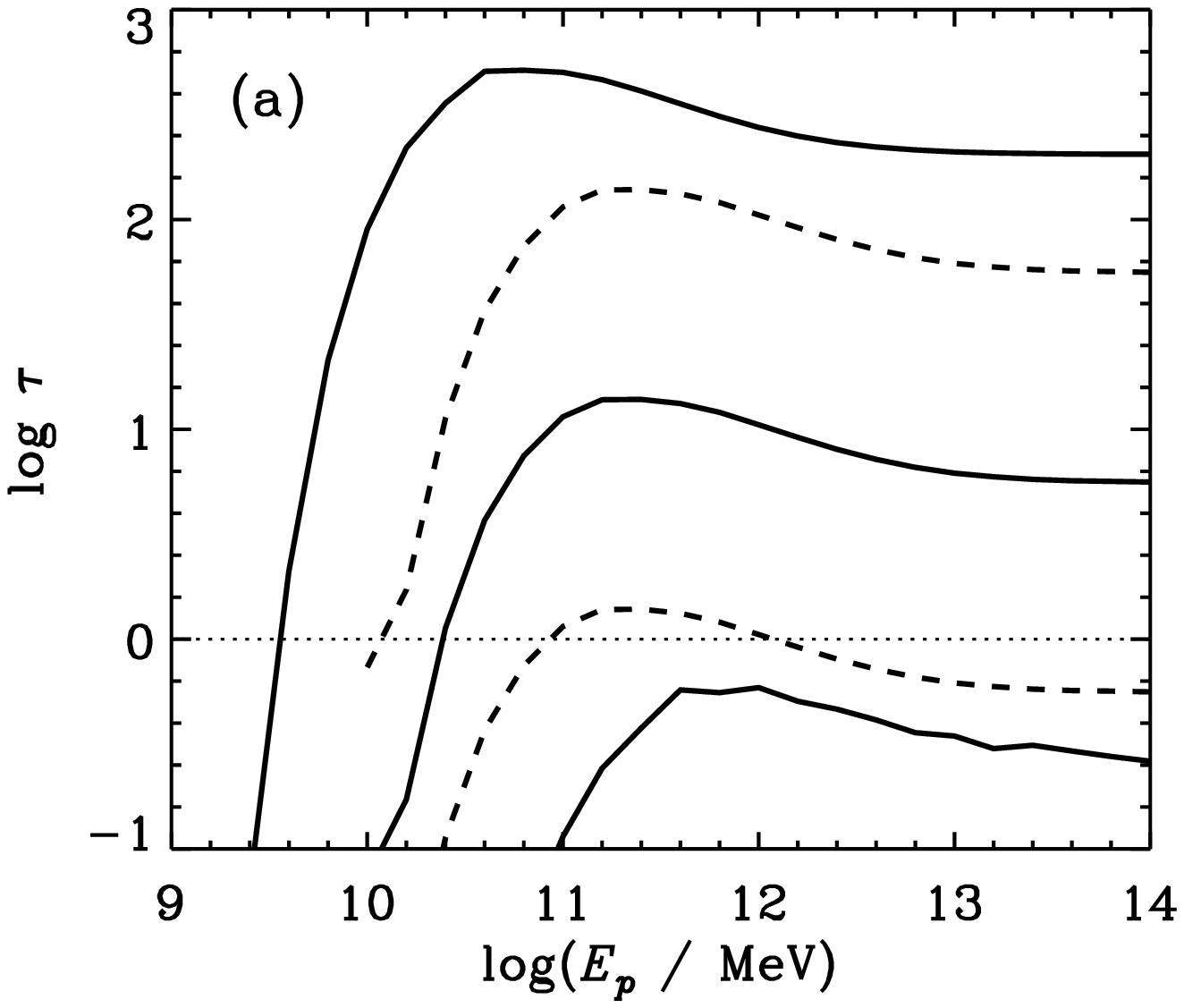}
\includegraphics{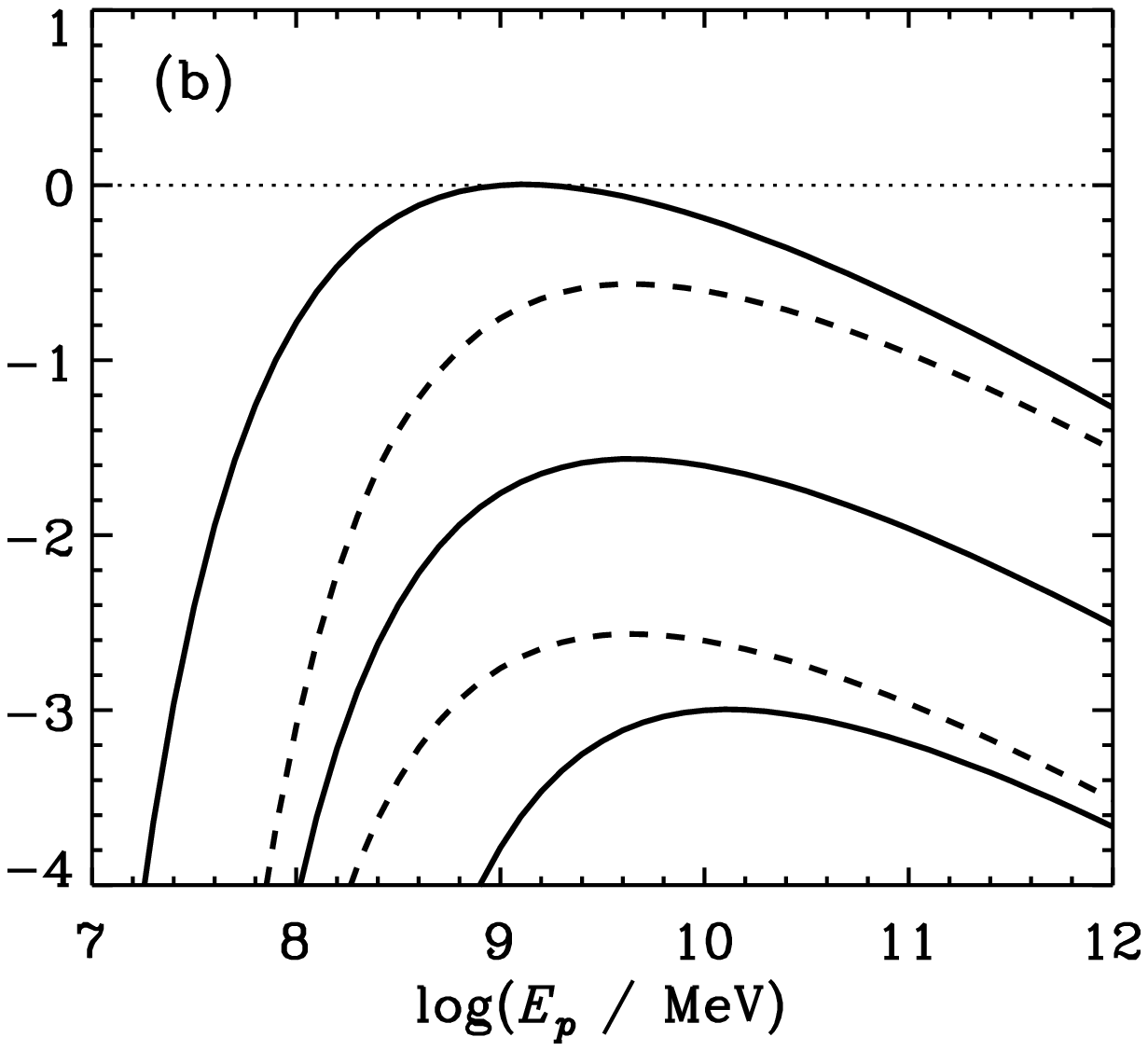}
      \caption[]{(a) Optical depth for pion production collisions
of protons with the disk photons; (b) optical depth for energy
loss of protons by production of $e^\pm$ pairs during
propagation. The full curves show the results for disk inner
radius $r_{\rm in} = 10^{15}$ cm and different inner disk
temperatures: $T_{\rm in} = 10^4$ (bottom), $3\times 10^4$, and
$10^5$ K (top); the dashed curves are for $T_{\rm in} = 3\times
10^4$ K and different inner disk radii: $r_{\rm in} = 10^{14}$
(bottom), and $10^{16}$ cm (top).}
\label{fig3}
    \end{figure*}

In order to simulate the propagation of protons and the
production of $\gamma$-rays and neutrinos, we first determine the
point of interaction of the proton with a soft photon coming from
the disk. The interaction length, $\lambda_{{\rm p}\gamma\pi}$,
for the proton in the disk radiation, as a function of its energy
$E_{\rm p}$ (Lorentz factor $\gamma_{\rm p}$ and velocity $\beta_{\rm
p}$) and its distance, $x$, from the disk surface is calculated,
by integrating the formula
\begin{eqnarray}
\lambda_{{\rm p}\gamma\pi}^{-1}(E_p, x) = \int \int n(\epsilon, \Omega)
\sigma_{{\rm p}\gamma\pi}(\epsilon') 
{{\epsilon'}\over{\epsilon}}d\epsilon d\Omega, 
\end{eqnarray}
\noindent
where $n(\epsilon, \Omega)$ is the differential density of
photons emitted by the disk, $\sigma_{{\rm
p}\gamma\pi}(\epsilon')$ is the energy dependent cross section
for pion production in proton--photon collision (Stecker~1968),
$d\Omega = d\phi d\mu$ is the solid angle of incident disk
photons, $\mu = \cos\alpha$ (see Fig.~\ref{fig1b}), $\epsilon' =
\epsilon \gamma_{\rm p}(1 + \beta_{\rm p}
\mu)$ and $\epsilon$ are energies of soft photons in the proton
and LAB frames, respectively.  To find out if protons have a
chance to interact with the disk photons for specific disk
parameters, the optical depth for protons is obtained by
integrating along the proton path
\begin{eqnarray}
\tau_{{\rm p}\gamma\pi} = 
\int_0^{\infty}\lambda_{{\rm p}\gamma\pi}^{-1}(E_{\rm p}, x) dx. 
\end{eqnarray}
\noindent
The optical depth for protons is shown in Fig.~\ref{fig3}a as a
function of proton energy, and for different parameters of the
disk radiation, $r_{\rm in}$ and $T_{\rm in}$.  It is clear that
for the disk parameters expected in AGNs protons have a
significant probability of interacting, and may even undergo
multiple interactions. In order to find out if the competing
process of $e^\pm$ pair production may become an important energy
loss mechanism for protons, we calculate the energy loss rates in
the disk radiation for $e^\pm$ pair production at a fixed
distance, $x$, from the disk surface
\begin{eqnarray}
{{dE_{\rm p}(x)}\over{dx}} = \int \int n(\epsilon, \Omega)
\sigma_{{\rm p}\gamma e^\pm}(\epsilon') K(\epsilon')
{{\epsilon'}\over{\epsilon}}d\epsilon d\Omega, 
\end{eqnarray}
\noindent
where the cross section for $e^\pm$ pair production,
$\sigma_{{\rm p}\gamma e^\pm}(\epsilon')$, and the inelasticity
$K(\epsilon')$ are taken from Chodorowski, Zdziarski \&
Sikora~(1992).  The optical depth for the total energy loss by
protons during their propagation through the disk radiation from
the disk surface to infinity is
\begin{eqnarray}
\tau_{{\rm p}\gamma e^\pm} = \int_0^{\infty}{{dE_{\rm p}(x)}\over{dx}}
{{1}\over{E_{\rm p}}} dx.
\end{eqnarray}
\noindent
This value is plotted in Fig.~\ref{fig3}b, for the same disk
parameters as used in Fig.~\ref{fig3}a. It is evident that proton
energy losses on $e^\pm$ pair production are not important except
for the case of very large disks with high inner disk
temperatures ($r_{\rm in} > 10^{15}$ cm and $T_{\rm in} > 10^5$
K). Such disks, with luminosities $L_{\rm disk} > 7\times
10^{46}$ erg s$^{-1}$, have not been observed.  The blazar 3C 273
has a disk luminosity a factor of about 3 lower, $L_{\rm
disk}\approx 2\times 10^{46}$ erg s$^{-1}$ (Lichti et al.~1995),
but this seems to be exceptionally high.  Therefore, we neglect
the influence of $e^{\pm}$ pair production on the propagation of
protons since these losses can be neglected in comparison to pion
production losses at energies above $E_{\rm p,min}$
(Eq.~\ref{eq5}). However this process can become important if
proton energies drop below the threshold for pion production far
from the accretion disk. Such pairs could contribute to the
formation of a broad synchrotron bump at low energies observed in
blazars.

If protons are very close to the disk, their interaction length
does not change significantly with distance $x_0$ from the disk.
Using the Monte Carlo method, the point of interaction of a
proton with the disk photon can be then determined from
\begin{eqnarray}
x_{\rm p} = x_0 - \lambda_{{\rm p}\gamma\pi} ln(1 - P_1),
\end{eqnarray}
\noindent
where $P_1$ is a random number. However at larger distances
from the disk the radiation becomes anisotropic, and the proton
interaction length can change significantly.  Then the distance
of proton interaction, $x_{\rm p}$, is obtained by solving
\begin{eqnarray}
P_1 = {\rm exp}[-\int_{x_0}^{x_{\rm p}}\lambda_{{\rm p}\gamma\pi}(l)^{-1}dl]
\end{eqnarray}
\noindent
where $l$ is the distance measured along the proton's path.  However,
if 
\begin{eqnarray}
P_1 < {\rm exp}[-\int_{x_0}^{\infty}\lambda_{{\rm p}\gamma\pi}(l)^{-1}dl]
\end{eqnarray}
\noindent the proton is assumed to have escaped from the disk
radiation.  The condition applied in order to distinguish between
quasi-isotropic and anisotropic cases is this same as used in
Bednarek (B97) (see footnote on page 5 in that paper).

Having selected randomly the distance, $x_{\rm p}$, at which
proton interacts with a disk photon, we have to sample the cosine
angle, $\mu$, of the photon from the disk which interacts with
the proton. The possible range of photon directions (from $\mu =
-1$ to $\mu_{\rm max}$, which corresponds to the outer disk
radius) is divided into one hundred intervals with width $\Delta
\mu$, corresponding to annular rings on the accretion disk
centered on $r_{\rm inj}$.  The ring defined by $\mu_{\rm
ph}(E_{\rm p},x_{\rm p})$, from which a disk photon is
interacting with the proton, is selected randomly by solving
\begin{eqnarray}
P_2 ={{\sum\limits_{-1}^{\mu_{\rm ph}(E_{\rm p},x_{\rm p})}\int 
\int n(\epsilon, \Omega)\sigma_{{\rm p}\gamma\pi}(\epsilon') 
{{\epsilon'}\over{\epsilon}}d\epsilon d\phi
\Delta\mu}\over{\sum\limits_{-1}^{\mu_{\rm max}}\int \int n(\epsilon, \Omega)
\sigma_{{\rm p}\gamma\pi}(\epsilon') 
{{\epsilon'}\over{\epsilon}}d\epsilon d\phi
\Delta\mu}}
\end{eqnarray}
\noindent
where $P_2$ is a random number.

Finally we have to sample the energy of the disk photon coming
from the ring $\mu_{\rm ph}$ for a proton located at a distance
$x_{\rm p}$. To save computing time we approximate the black
body spectrum of photons emitted at distance $r$ from the disk
center by monoenergetic photons with energies corresponding to
the mean photon energy of the black body spectrum, $\epsilon =2.7
k_{\rm B} T(r)$, where $k_{\rm B}$ is the Boltzman constant, and
$T(r) = T_{\rm in} (r/r_{\rm in})^{-3/4}$, as predicted by
Shakura \& Sunyaev model. In such a case, the energy of photons
coming from a specific ring $\mu_{\rm ph}$ is uniquely defined by
the azimuthal angle $\phi$ (see Fig.~\ref{fig1b}). We can sample
the angle $\phi$ randomly by solving
\begin{eqnarray}
P_3 = {{\int\limits_{\phi_{\rm min}}^{\phi_{\epsilon}} 
n(\epsilon(\phi),\mu) \sigma_{{\rm p}\gamma\pi}(\epsilon') 
{{\epsilon'}\over{\epsilon}}d\phi}
\over{\int\limits_{\phi_{\rm min}}^{\pi} 
n(\epsilon(\phi),\mu) \sigma_{p\gamma\pi}
(\epsilon') {{\epsilon'}\over{\epsilon}}d\phi}}
\end{eqnarray}
\noindent
where $P_3$ is a random number and $\phi_{\rm min}$ is equal to
0, or is calculated for the given inner disk radius $r_{\rm in}$
if the ring defined by $\mu$ lies partially inside $r_{\rm in}$.  
The interacting photon is now completely specified,
allowing us to compute its energy in the proton frame $\epsilon'
= \epsilon(\phi_{\epsilon})\gamma_{\rm p} (1 + \beta_{\rm p}\mu_{\rm
ph})$, model the interaction, and obtain the energies of pions
produced in the proton frame.
 
The above procedure is repeated until the relativistic proton
escapes from the disk radiation. The parameters of produced pions
are stored for further computation of neutrino and $\gamma$-ray
spectra.

\subsection{Spectra of neutrinos and gamma-rays from decay of pions}

Because of the complexity of the present model, in particular the
cascading in the radiation and magnetic fields, we have had to
make a number of approximations in order to save computer time.
In particular, in the calculation of spectra of secondary
particles produced in interactions of protons with soft photons
was, of necessity, subject to approximation.  However, we include
single pion production (in the $\Delta$ resonance region)
although in an approximate way, and the effects of multiple pion
production far away from the threshold is treated using an
average multiplicity relation similar to that found in low energy
$pp$ interactions,
\begin{eqnarray}
<n_\pi> \propto s^{0.25},
\end{eqnarray}
\noindent
where $s = m_{\rm p}^2 + 2\epsilon E_{\rm p} (1 +\beta_{\rm p}\mu)$ is 
the center of momentum frame energy squared (in this section we define $c
\equiv 1$), and $m_{\rm p}$ is the proton rest mass. The ratio of charged
pions to all pions produced in a single multiple pion interaction
is taken to be $n_{\pi^{\pm}}/n_{\pi} \approx 2/3$.

In the case of single pion production ($\Delta$ resonance decay)
the pion energies in the center of momentum system (CMS) are
given by,
\begin{eqnarray}
E_\pi^{\rm CMS} ={{2m_{\rm p} \epsilon' + m_\pi^2}\over{2\sqrt{s}}}.
\end{eqnarray}
\noindent 
It is assumed that the $\Delta$ resonance decays isotropically,
so that pions are produced isotropically in the CMS and we
may sample their direction cosine angle using
\begin{eqnarray}
\mu_\pi = 2 P_4 - 1,
\end{eqnarray}
\noindent
where $P_4$ is a random number. The energy of the pion in the
LAB frame is then
\begin{eqnarray}
E_\pi^{\rm LAB} = \gamma_{\rm CMS} (E_\pi^{\rm CMS} + \beta_{\rm CMS} 
p_\pi^{\rm CMS} \mu_\pi),
\end{eqnarray}
\noindent
where $\gamma_{\rm CMS} = (E_p+\epsilon)/\sqrt{s}$, $\beta_{\rm
CMS}=\sqrt{1 - 1/\gamma_{\rm CMS}^2}$, and $p_\pi^{\rm CMS}$
is the pion momentum in the CMS.

   \begin{figure*}
      \vspace{7.cm}
\includegraphics{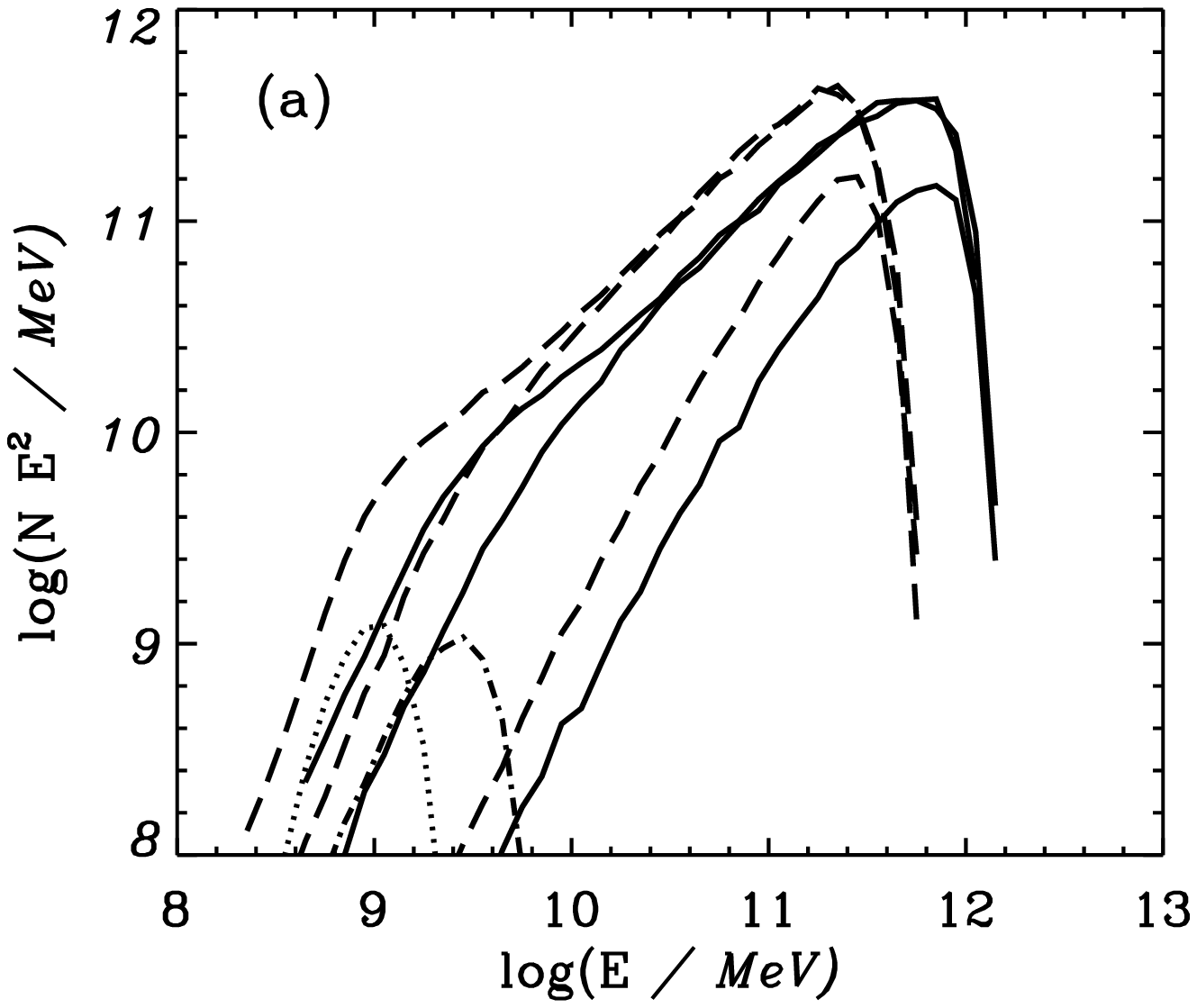}
\includegraphics{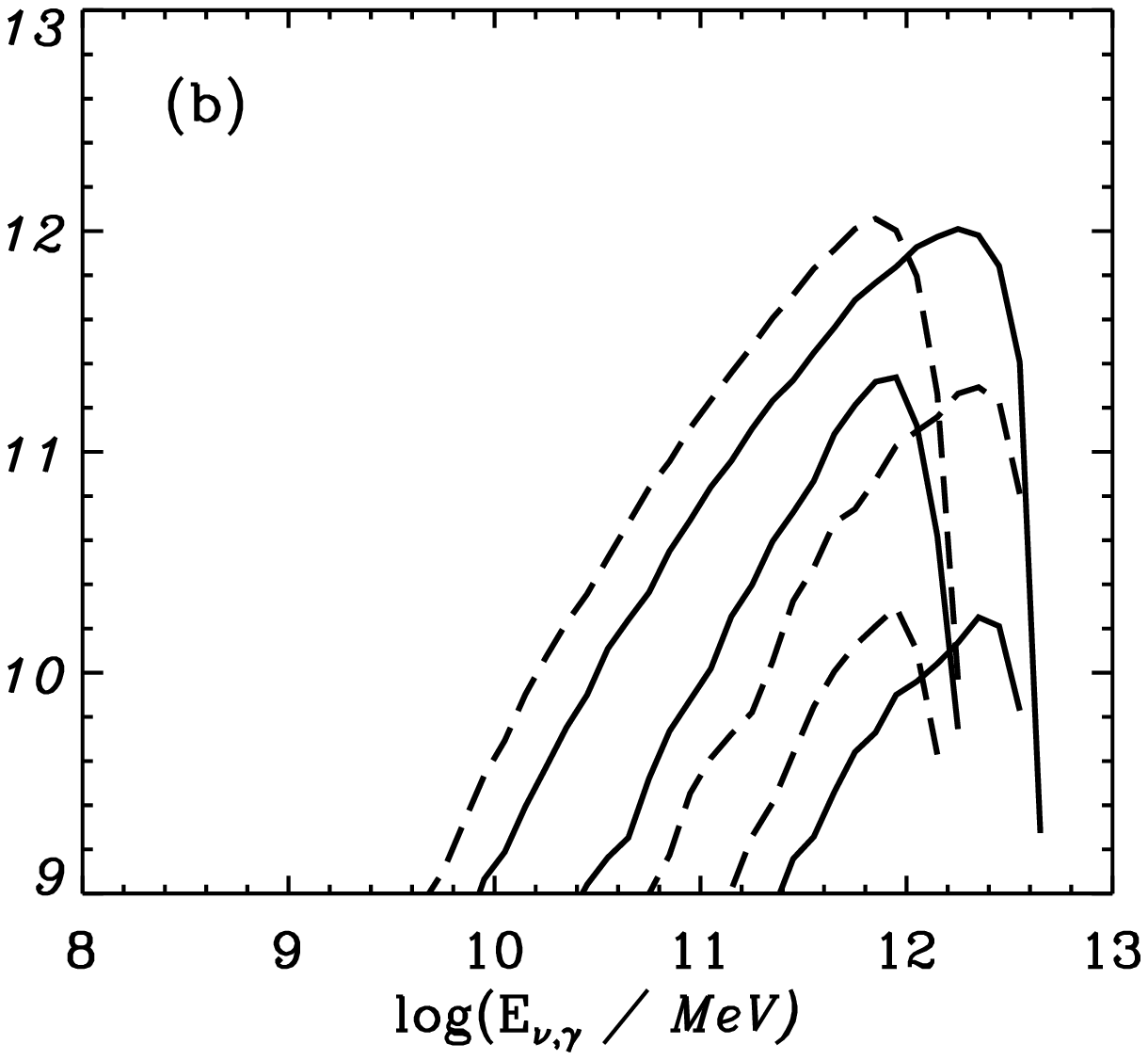}
      \caption[]{The spectra of $\gamma$-rays (full curves) and
muon-neutrinos (long-dashed curves) from decay of pions which are
produced in collisions of protons with the disk photons. The
parameters used are: (a) $T_{\rm in} =3\times 10^4$ K, $l_{rec} =
10^{12}$ cm (giving $E_p = 3\times 10^{12}$ MeV); (b) $T_{\rm in}
= 10^4$ K, $l_{rec} = 3\times 10^{13}$ cm ($E_p = 10^{13}$ MeV).
In each case results are given for $r_{\rm in} =10^{14}$ cm
(bottom), $10^{15}$ cm, and $10^{16}$ cm (top). The dotted and
dot-dashed curves in (a) show the spectra of neutrinos and
$\gamma$-rays for protons with $E_p = 3\times 10^{10}$ MeV.  }
\label{fig4}
    \end{figure*}

The pions decay isotropically in their rest frame producing:
$\gamma$-rays, muon neutrinos, and muons with energies which we
approximate by
\begin{eqnarray}
E_\gamma^\pi = 0.5 m_\pi,
\end{eqnarray}
\begin{eqnarray} 
E_\nu^\pi = 0.5 (m_\pi^2 - m_\mu^2)/m_\pi, 
\end{eqnarray}
and 
\begin{eqnarray}
E_\mu^\pi = 0.5 (m_\pi^2 + m_\mu^2)/m_\pi, 
\end{eqnarray}
where $m_\pi$ is the rest mass of a pion (we neglect the mass
difference between neutral and charged pions).  The cosine angles
of decay products of pions can be sampled randomly from
\begin{eqnarray}
\mu_{\gamma,\nu}^\pi = 2 P_5 - 1,
\end{eqnarray}
\noindent
where $P_5$ is a random number, and $\mu_\mu = -\mu_\nu$.
Their energies in the LAB frame are given by
\begin{eqnarray}
E_{\gamma,\nu, \mu}^{\rm LAB} \approx 
E_{\gamma,\nu,\mu}^\pi \gamma_\pi (1 + \beta_\pi \mu_{\gamma,\nu,\mu}),
\end{eqnarray}
\noindent
where $\gamma_\pi = E_\pi^{\rm LAB}/m_\pi$, and $\beta_\pi =
\sqrt{1 - 1/\gamma_\pi^2}$. The energies of muon-neutrinos from
the decay of secondary muons are simply estimated by taking
$E_{\nu}^{\mu}\approx E_{\mu}/3$.

The spectra of muon-neutrinos and $\gamma$-rays, averaged over
the propagation of protons along the direction perpendicular to
the disk surface, are shown in Fig.~4. Fig.~\ref{fig4}a shows the
spectra for the local disk temperature in the region of magnetic
reconnection $T = 3\times 10^4$K, $l_{\rm rec} = 10^{12}$ cm, and
$r_{\rm inj} = 1.5r_{\rm in}$. For these parameters, protons are
injected from the reconnection region with energies $E_{\rm p} =
5\times 10^{12}$ MeV (see Eq.~\ref{eq8}).  The full curves show
the spectra of $\gamma$-rays for the inner disk radii: $r_{\rm
in} = 10^{14}, 10^{15}$, and $10^{16}$ cm.  The corresponding
neutrino spectra are shown by the dashed curves.  The spectra of
$\gamma$-rays and neutrinos in the case of protons injected with
energies $E_{\rm p} = 3\times 10^{10}$ MeV are shown by the
dot-dashed and dotted curves, respectively.  In Fig.~\ref{fig4}b
we show the spectra of neutrinos and $\gamma$-rays for $T = 10^4$
K, $l_{\rm rec} = 3\times 10^{13}$ cm, $r_{\rm inj} = 1.5r_{\rm
in}$. For these parameters the proton initial energy is $E_{\rm
p} = 10^{13}$ MeV.

\section{Cascade initiated by secondary gamma-rays}

As shown in Figs.~4, the $\gamma$-rays originating from decay of
neutral pions have extremely high energies. Depending on the
conditions above the reconnection region, they can develop
cascades in the perpendicular component of the magnetic field via
magnetic $e^{\pm}$ pair production, and subsequent emission of
synchrotron photons, provided that their energies fulfill the
condition (Erber~1966, B97)
\begin{eqnarray}
\chi \equiv E_\gamma B_\perp (m_{\rm e} c^2 B_{\rm cr})^{-1} \ge 0.05,
\end{eqnarray}
\noindent
where $B_\perp$ is the component of the magnetic field which is
perpendicular to the photon propagation, $E_\gamma$ is the photon
energy, $B_{\rm cr}\equiv m_{\rm e}^2 c^3 (e h)^{-1} =
4.414\times 10^{13}$ G is the critical magnetic field strength,
and $m_{\rm e}$ is the electron rest mass.

For hot disks having weak magnetic fields, the $e^\pm$ pair
energy losses by ICS dominate over synchrotron losses, and the
secondary $e^{\pm}$ pairs initiate ICS $e^{\pm}$ pair cascades
(e.g. Protheroe, Mastichiadis \& Dermer~1992, Coppi, Kartje \&
K\"onigl~1993). In the present work we discuss the case where the
magnetic and disk radiation fields are initially in equipartition
close to the surface. Therefore the magnetic energy density
dominates over the radiation energy density above the disk since
the magnetic field drops more slowly with the distance from the
disk than the radiation field (Bednarek, Kirk \&
Mastichiadis~1996b).

   \begin{figure*}
      \vspace{7.cm}
\includegraphics{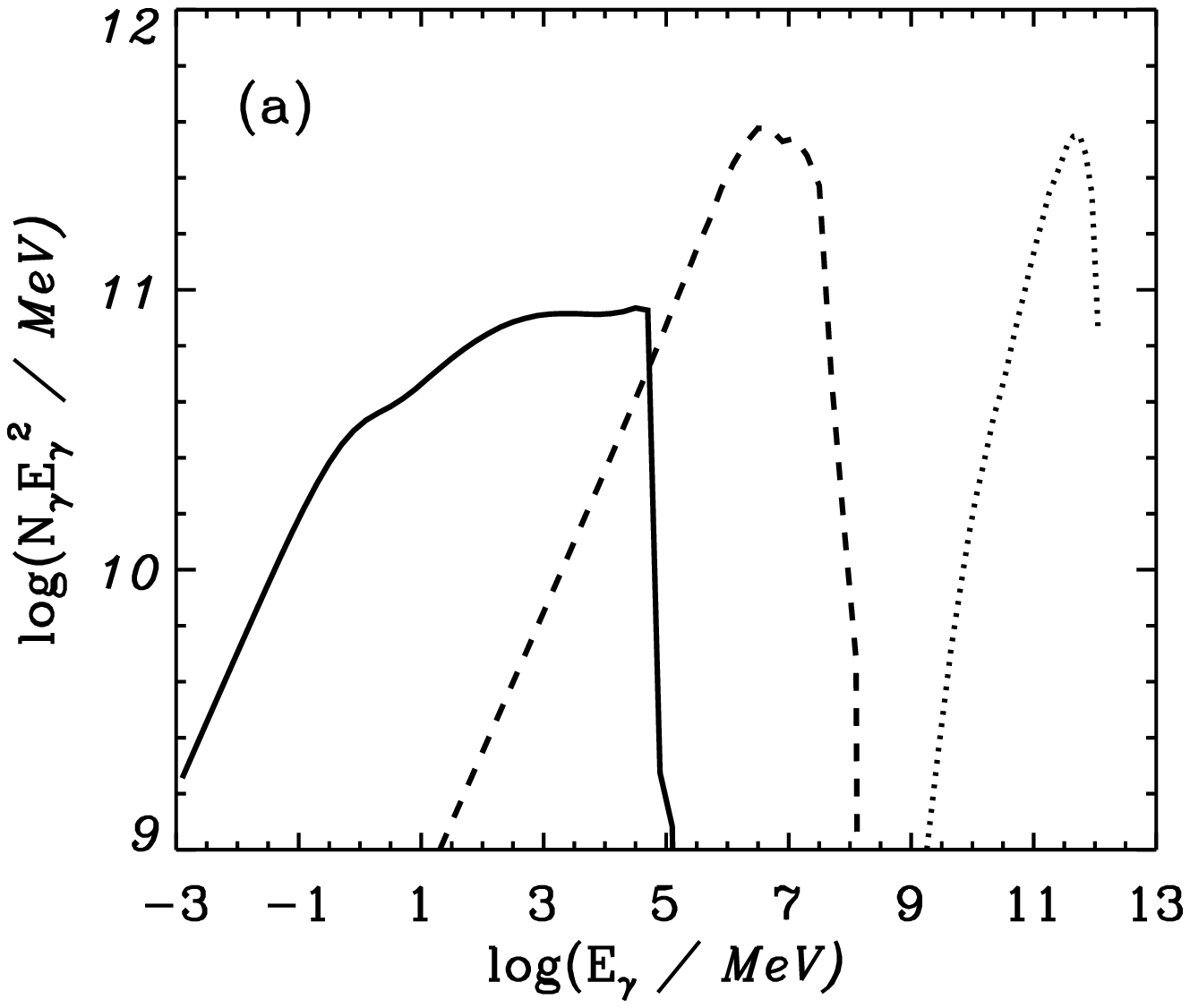}
\includegraphics{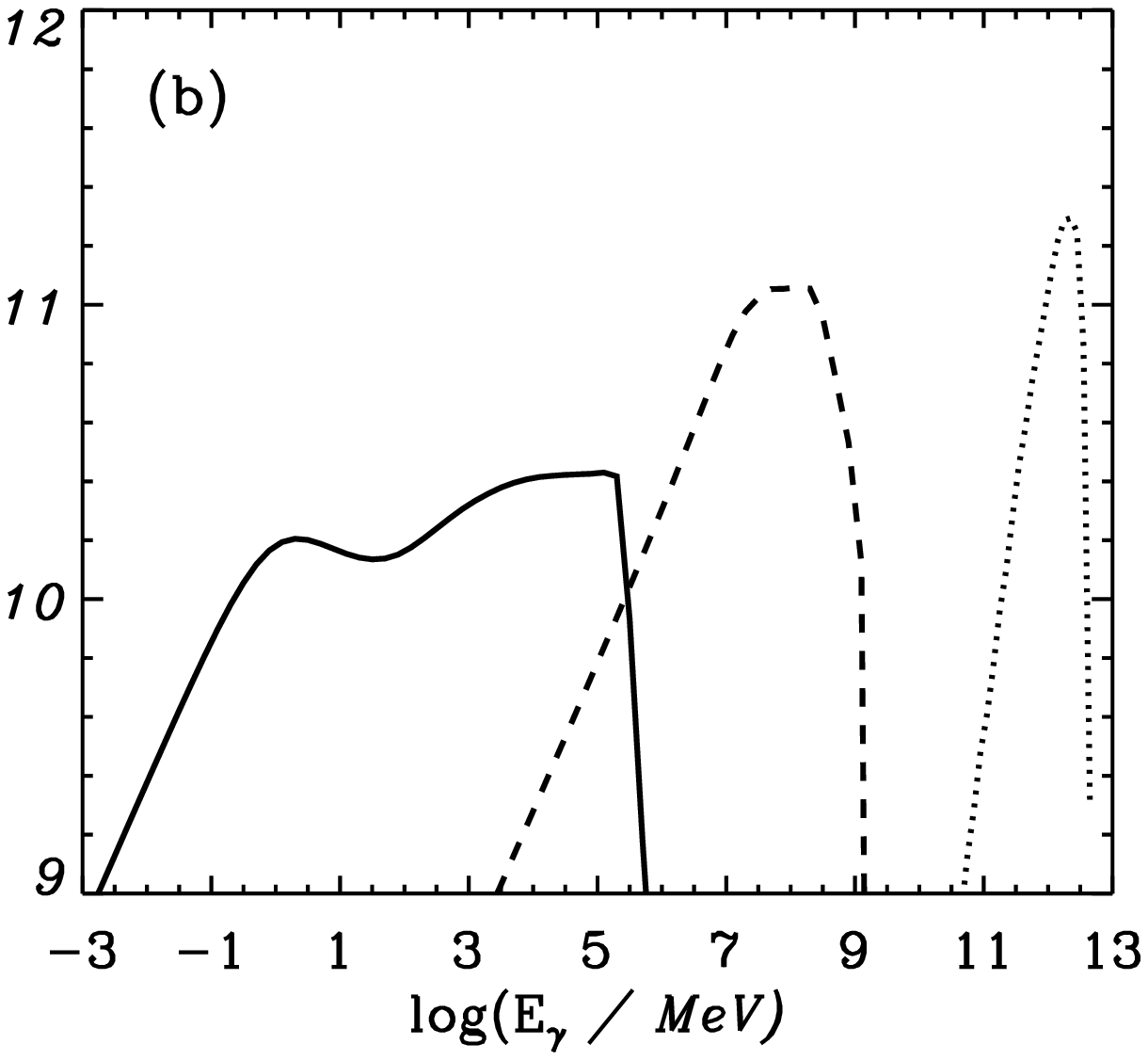}
      \caption[]{The $\gamma$-ray spectra obtained in a sequence
of processes, discussed in the text, which are initiated by
primary $\gamma$-rays produced for the parameters of the model as
in Figs.~4(a) and 4(b), respectively, and $r_{\rm in} =10^{15}$ cm.  The
$\gamma$-ray spectra from decay of neutral pions produced by
protons in collisions with the disk photons are shown by the
dotted curves; the $\gamma$-ray spectra emerging from the cascade
in the magnetic field above the disk are shown by the dashed
curves; and the $\gamma$-ray spectra from absorption of the
cascade $\gamma$-ray spectra in the disk radiation, and
subsequent synchrotron cooling of secondary $e^\pm$ pairs are
shown by the full curves.}
\label{fig5}
    \end{figure*}

The energies of $\gamma$-ray photons produced by protons and
their distances from the disk on production are stored and used
as input to the subsequent cascading in the magnetic field
calculated by using the numerical cascade code described in
(B97). A sequence of different processes is involved
during propagation of these $\gamma$-rays.  First, the
$\gamma$-rays cascade in the perpendicular component of the
magnetic field above the disk, and we compute the spectra of
secondary $\gamma$-rays and $e^{\pm}$ pairs emerging from this
synchrotron $e^\pm$ cascade.  At distances further from the disk,
where the cascade in the magnetic field becomes inefficient, the
propagation of $\gamma$-rays in the disk radiation is
considered. If the inner disk temperature is high enough, some of
secondary $\gamma$-rays are absorbed creating a second generation
of $e^{\pm}$ pairs.  These pairs cool predominantly in the local
magnetic field by the synchrotron process, emitting a second
generation of $\gamma$-rays. Some of the most energetic
$\gamma$-rays from second generation can further be absorbed in
the disk radiation producing tertiary $e^{\pm}$ pairs which cool
by the synchrotron process as well. Since the optical depth for
tertiary $\gamma$-rays in the disk radiation is low, they usually
escape unabsorbed. The emerging $\gamma$-ray spectrum is obtained
by summing the contributions from all these processes described
above.  In the present work we use a more general prescription
for the dependence of the perpendicular component of the magnetic
field along the jet ($B_{\perp}$) than applied by
Bednarek~(B97). Here we use
\begin{eqnarray}
B_{\perp}(x) = B_0 (x/x_1)^{-\beta}.
\end{eqnarray}
\noindent
which for the cases with $\beta < 1.5$ guarantees that the
magnetic energy density always dominates over the disk, radiation
energy density above the disk provided that it dominates at the
base of the jet~(Bednarek, Kirk \& Mastichiadis~1996b).
 
The spectra of $\gamma$-rays from the synchrotron $e^\pm$
cascade, and the final $\gamma$-ray spectra escaping from the
radiation field of the accretion disk, are shown in Fig.~5 by
the dashed and full curves, respectively. They correspond to the
initial $\gamma$-ray spectra from the decay of pions shown in
Fig. 4, and for a disk inner radius $r_{\rm in} = 10^{15}$ cm. In
these simulations, we apply the following parameters which define
the structure of the magnetic field in the jet above the
reconnection region: $B_0$ is taken to be equal to the value of
the magnetic field close to the disk surface, $x_1 = r_{\rm
inj}$, and $\beta = 1.25$. Our simulations show that the shape of
the final $\gamma$-ray spectrum is not very sensitive to $\beta$
for values in the range between 1. and 1.5.

\section{June 1991 gamma-ray flare from 3C 279 and expected neutrinos}

The $\gamma$-ray spectra produced in the sequence of processes
discussed above are consistent with the spectra observed from
FSRQ blazars. As an example, in Fig.~\ref{fig6} we compare the
$\gamma$-ray spectrum, shown in Fig.~\ref{fig5}a, with the
observations of the flare from 3C 279 during June 1991 by the
COMPTEL, EGRET and Ginga detectors (Williams et al.~1995, Kniffen
et al.~1993, Hartman et al.~1996). The accompanying neutrino
flare (marked by the dashed curve) corresponds to $\sim 28$ muon
neutrinos per day per km$^2$ with a typical energy of $\sim
10^8$ GeV. Such a flux, although small, may still be detectable
since it is many orders of magnitudes above the neutrino
background induced in the atmosphere within the $1^\circ$
(Lipari~1993) shown by the dot-dashed curve in Fig.~\ref{fig6}.
The detection probability of neutrino telescopes operating at
this energy can be estimated to be $\sim 3\times 10^{-3}$
(Gaisser, Halzen \& Stanev~1995), giving a rate of $\sim 0.1$
muons per day per km$^2$. Therefore, for typical flares occurring
on a time scale of a week one coincident neutrino event should be
detected provided that the neutrino is massless. During a period
of years, a large neutrino detector should be able to
definitively answer the question of the importance of hadronic
processes by measurements of correlations between neutrino
arrival directions and directions to known FSRQ blazars.

The spectrum of neutrinos expected in this model is limited to a
relatively narrow energy range if compared to other hadronic
models (e.g. Protheroe~1997, Nellen, Mannheim \& Biermann~1993,
Mannheim~1995, and comparison of different models in
Protheroe~1998).  The reason is that we have considered the case
in which all hadrons are accelerated in the reconnection region
to this same maximum energy, in contrast to other models
mentioned in the Section~1, which assumed the injection of a
power-law spectrum of protons.  Therefore, the shape of the
neutrino spectrum may also give us important information on the
acceleration mechanism of hadrons in blazars.

\section{Discussion and Conclusion}

\begin{figure}
      \vspace{7.cm}
\includegraphics{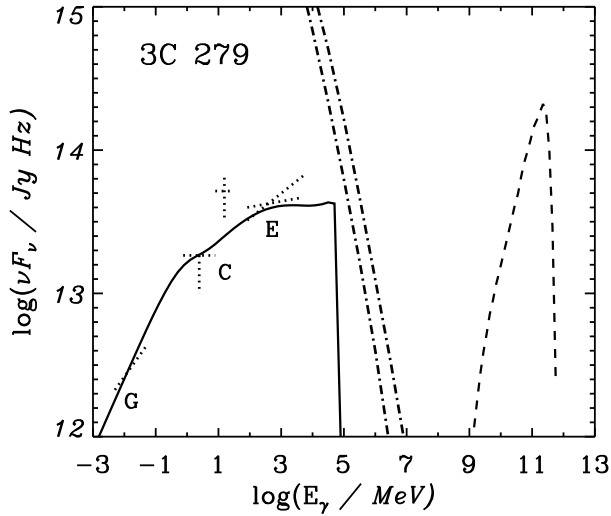}
      \caption[]{The $\gamma$-ray spectrum obtained from the
present model (full curve) is compared with observations of
the June 1991 flare from 3C 279 detected by COMPTEL
(C) (Williams et al.~1995), EGRET (E) (Kniffen et al.~1993) and Ginga
(G) (Hartman et al.~1996).  The parameters of the fit are as in
Fig.~\ref{fig5}a. The dashed curve shows the expected muon
neutrino flux during this flare. The dot-dashed curves show the
atmospheric neutrino background (horizontal - upper curve, and
vertical - lower curve) for a neutrino detector with $1^\circ$
angular resolution .}
\label{fig6}
    \end{figure}

Some models of magnetized accretion disks predict the appearance
of relatively small, but efficient, acceleration regions close to
the surface of the inner part of the accretion disk as a result
of magnetic reconnection (Haswell, Tajima \& Sakai~1992, Lesch \&
Pohl~1992). In this paper, we discuss the consequences of
acceleration of hadrons in such regions. Depending on the
conditions in the acceleration region, and above the disk (see
Fig.~\ref{fig1}), accelerated hadrons may either escape, initiate
cascades after emerging from the acceleration region, or produce
copious secondary leptons during the acceleration process which
initiate subsequent leptonic cascading.  Primary leptons can also
initiate cascades, and this possibility has been already
discussed~(B97). In the analysis of hadronic cascades
we included the processes of pion production by hadrons,
synchrotron radiation of secondary $e^\pm$ pairs in the quantum
domain, magnetic $e^\pm$ pair production by $\gamma$-ray photons
and absorption of the $\gamma$-rays in the disk radiation. The
$\gamma$-ray spectra emerging from such a sequence of processes
resemble the spectra observed from FSRQ blazars (Figs~5).  Their
comparison with the June 1991 flare from 3C 279 allows one to
predict the strength of the accompanying neutrino flare (see
Fig.~\ref{fig6}).

In the present model we have neglected the contribution of
interactions of accelerated hadrons with the background matter,
either during the acceleration stage, or after propagation
through the disk radiation.  Interactions of relativistic protons
with the disk radiation dominate over the interactions with
matter if the density of matter $n_{\rm H}$ is lower than
\begin{eqnarray}
n_{\rm H}\approx 2\times 10^{11} T_4^3 ~{\rm particles~ cm^{-3}},
\end{eqnarray}
\noindent
If the density of matter is relatively high at larger distances
from the disk, then protons, at this stage having lost most of
their energy during propagation through the disk radiation, can
also produce neutrinos by interactions with matter. These
neutrinos are typically of lower energies, and their power will
be much less than in the higher energy neutrinos from
interactions with radiation.

Here we have considered only a single acceleration region in
order not to complicate the picture too much. However, in
principle many acceleration regions with different parameters
could be present simultaneously on the disk surface. Different
types of cascades may then contribute to the observed
$\gamma$-ray spectrum from a specific blazar, and to the neutrino
flux. If multiple reconnection sites do contribute, then the
neutrino fluxes predicted for a single site should be considered
as upper limits. However, a flare being produced by a single
reconnection region seems to be equally probable given the likely
short duration of each reconnection region ($\sim l_{\rm
rec}/c$).

It is expected that as a result of magnetic reconnection a large
amount of energetic plasma may enter the jet. Therefore, the
cascade processes taking place above the reconnection region
should also occur in the jet plasma which moves with mildly
relativistic speeds. Hence, flares of the type discussed above
may be accompanied by additional radiation processes caused by
this energetic plasma. Recent multiwavelength observations of Mrk
501 argue that high energy processes in blazars are complicated
(see e.g. Pian et al.~1998, Protheroe et al.~1998), and more than
one different mechanism may be responsible for producing the
observed photon spectrum.

The motion of plasma, and additionally the longitudinal component
of the magnetic field in the jet (we do not include its influence
on the final $\gamma$-ray spectrum) could provide collimation of
the produced radiation in the direction perpendicular to the disk
surface.

In conclusion, this model predicts that two types of high energy
flares may occur if the particles (hadrons, leptons) are
accelerated to high energies in reconnection regions on the
accretion disk surface.  Pure $\gamma$-ray flares can be caused
either by secondary $e^\pm$ pairs produced during acceleration of
protons which collide with the soft disk photons, or by primary
leptons accelerated in the reconnection region.  In this latter
case, the $\gamma$-ray spectrum can extend up to TeV energies if
the disk temperature is low enough. For large, luminous disks,
the magnetic field energy can be efficiently transferred to
relativistic protons which can initiate $\gamma$-ray and neutrino
flares via production of pions in collisions with soft disk
photons.  Therefore, we suggest that blazars of FSRQ type, for
which $\gamma$-ray spectra show some evidence of a cut-offs at
higher energies (Pohl et al.~1997) and features of large
accretion disks (emission lines), may become sources of GeV
$\gamma$-rays and $\sim 10^{17}$ eV neutrinos.  In contrast, BL
Lac objects, with spectra extending to TeV energies, should not
emit high energy neutrinos according to the present model.

\section*{acknowledgments}

W.B. thanks the Department of Physics and Mathematical Physics at
the University of Adelaide for hospitality during his visit.
This research is supported by a grant from the Australian
Research Council. The work of W.B. is partially supported by the Polish
{\it Komitet Bada\'n Naukowych} grant 2 P03D 001 14.

\end{document}